\documentstyle[amstex,aps]{revtex}

\begin{document}
\title{Growth-Collapse Cycles of a Bose-Einstein Condensate with Attractive
Interactions}
\author{Kerson Huang}
\address{Center for Theoretical Physics,\\
Massachusetts Institute for Technology,\\
Cambridge, MA, 02139, USA}
\date{\today }
\maketitle

\begin{abstract}
A Bose-Einstein condensate of atoms with attractive interactions exhibits
growth and collapse cycles, when it is fed by a thermal cloud. Recently this
phenomenon has been directly observed in a trapped $^{7}$Li gas. We offer a
quantitative explanation of the data, on the basis of a model proposed
earlier. It is shown that the condensate wave function acquires a chaotic
component after the first collapse, indicating superfluid turbulence.
\end{abstract}

\section{Introduction}

In an elegant experiment \cite{Gerton}, Hulet and his team at Rice
University have observed the growth and collapse of \ a Bose-Einstein
condensate of magnetically trapped $^{7}$Li atoms, whose interactions are
predominantly attractive.\ This fascinating system had been created \cite%
{Bradley} and studied \cite{Sackett} by the same group some time ago; but
the instability of the system has not been directly observed until now. The
purpose of this paper is to give a quantitative explanation of the data,
based on a theoretical model proposed earlier\cite{UedaHuang}\cite%
{Eleftheriou}. We refer to these references for earlier literature.

Because of the attractive interactions,$^{7}$Li atoms in free space would
solidify at low temperatures. When confined in an external potential,
however, the atoms are maintained in a gaseous state by their zero-point
motion, as long as their number is small. Above a critical number, however,
the attractive interactions overcome the kinetic energy, and the system
collapses into a state of high density. This instability is similar to that
of a white dwarf star whose mass is greater than the Chandrasekhar limit, in
which the gravitational attraction overwhelms the zero-point pressure of the
Fermi gas of electrons in the star. Unlike in a white dwarf star, however,
the particle number here is not conserved, for the atoms can be expelled
from the magnetic trap due to two- and three-body collisions, which become
increasingly important at higher densities. During the collapse, the
particle number suddenly drops to a subcritical value, only to grow back
subsequently, if the condensate is fed by a surrounding thermal cloud. Thus,
the condensate undergoes cycles of growth and collapse, and in this respect
the system is more akin to an electric buzzer than a white dwarf star.

Let us first ignore the gain and loss mechanisms. To model the system, we
treat the condensate in the mean-field approximation, an assume that the
condensate wave function $\psi $ satisfies a Gross-Pitaevskii equation, or
nonlinear Schr\"{o}dinger equation (NLSE):

\begin{eqnarray}
i\hbar \frac{\partial \psi }{\partial t} &=&\left[ -\frac{\hbar ^{2}}{2m}%
\nabla ^{2}+V(r)-U_{0}|\psi |^{2}\right] \psi ,  \nonumber \\
U_{0} &=&\frac{4\pi \hbar ^{2}|a|}{m},  \label{NLSE}
\end{eqnarray}%
where $a$ is a negative scattering length, and the external potential is
taken to be harmonic: 
\begin{equation}
V(r)=\frac{1}{2}m\omega ^{2}r^{2}.
\end{equation}%
This defines a characteristic length $d_{0}$, the width of the unperturbed
ground-state wave function: 
\begin{equation}
d_{0}=\sqrt{\frac{\hbar }{m\omega }}.
\end{equation}%
The number of condensate particles enters through the normalization 
\begin{equation}
N=\int d^{3}r|\psi |^{2}.
\end{equation}%
The corresponding Hamiltonian is given by%
\begin{equation}
H=\int d^{3}r\left[ -\frac{\hbar ^{2}}{2m}\psi ^{\ast }\nabla ^{2}\psi
+V(r)\psi ^{\ast }\psi -\frac{U_{0}}{2}(\psi ^{\ast }\psi )^{2}\right] .
\label{Hamiltonian}
\end{equation}%
In the Rice experiment, the parameters have the values

\begin{align}
a& =-1.45\text{ nm,}  \nonumber \\
d_{0}& \approx \text{3.16 }\mu \text{m,}  \nonumber \\
\omega & \approx 908\text{ s}^{-1}.  \label{parameters}
\end{align}%
Actually, the magnetic trap was not quite spherically symmetric, and $\omega 
$ here represents the geometric mean of the frequencies.

\ In the absence of an external potential in spatial dimensions $D\geq 2$,
the NLSE with attractive interactions ($U_{0}>0)$ exhibits\ ``self-focusing''%
\cite{Sulem}, in which the system collapses to a state of locally infinite
density in a finite time. The cause of this instability is the term $-\frac{1%
}{2}U_{0}|\psi |^{2}$ in the Hamiltonian \ref{Hamiltonian}, which has no
lower bound as the density $|\psi |^{2}$ increases. In the presence of an
external trap $V(r)$, however, there exists an energy barrier against
collapse. As a function of $|\psi |^{2}$, the form of this barrier depends
on the location $r$, and its height is smallest at $r=0$, where the external
potential is weakest. As predicted in \cite{UedaHuang}, and verified through
numerical calculations in \cite{Eleftheriou}, the energy barrier at $r=0$ is
breached when $N>N_{c}$, with%
\begin{equation}
N_{c}=0.574\frac{d_{0}}{|a|}=1250,
\end{equation}%
and this triggers the collapse of the system. At the center of the trap,
there appears a black hole into which particles are drawn, and they will be
drained from the system when loss mechanisms are taken into account. Of
course, particles can surmount the energy barrier through quantum tunneling;
even when $N<N_{c}$. but this is a slow process that can be neglected
compared to loss through collisions.

What happens to the particles that go through the energy barrier? According
to the Hamiltonian \ref{Hamiltonian} their density would continue to
increase indefinitely; but, of course, once the density begins to grow,
effects so far neglected will become important. An example is
solidification, which might be simulated by adding high-order terms in the
Hamiltonian, such as $c|\psi |^{6}$, with $c>0$. Such a term in the NLSE
will lead to the formation of droplets in a first-order phase transition %
\cite{Josserand}. However, such effects are overshadowed by loss though
collisions, and will be ignored.

\section{Gain and loss mechanisms}

Ketterle and his team at MIT has measured the rate of growth of a $^{23}$Na
condensate fed by a thermal cloud, and fitted the results to the following
formula \cite{Miesner}:%
\begin{equation}
\frac{dN}{dt}=c_{0}N\left[ 1-\left( \frac{N}{N_{\text{eq}}}\right) ^{0.4}%
\right] ,
\end{equation}%
with $c_{0}=(50$ ms)$^{-1},N_{\text{eq}}\approx 5\times 10^{5}$. The factor
in square brackes represents a saturation effect. We shall ignore it by
assuming $N<<N_{\text{eq}}$ for our case. The growth rate $c_{0}$ should be
proportional to the elastic scattering cross section, and hence to the
square of the scattering length. Since the scattering lengths for $^{23}$Na
and $^{7}$Li are $2.75$ nm and $-1.45$ nm respectively, we take for $^{7}$Li 
\begin{equation}
c_{0}=\left( \frac{1.45}{2.75}\right) ^{2}(50\text{ ms})^{-1}.
\end{equation}

The loss rate $R(N)$ due to two and three-body collisions have been
calculated by Dodd {\it et al \cite{Dodd}, }who report their results in the
form%
\begin{eqnarray}
R(N) &=&A\int d^{3}r|\psi |^{4}+B\int d^{3}r|\psi |^{6},  \nonumber \\
A &=&1.2\times 10^{-14}\text{ cm}^{3}\text{ s}^{-1},  \nonumber \\
B &=&2.6\times 10^{-28}\text{ cm}^{6}\text{ s}^{-1},
\end{eqnarray}%
where $A$ is due to two-body collisions, and $B$ is due to three-body
collisions. The experimental value of the two-body loss rate $A$ is
consistent with the value give above, provided it is interpreted as the loss
rate per event \cite{Gerton1}. In view of this fact, we reinterpret both $A$
and $B$ as rates {\it per event} rather than per particle. Since 2 particles
are lost in a two-body collision, and 3 particles are lost in a three-body
collision, the loss rates per particle is taken to be $2A$ for two-body
collisions, and $3B$ for three-particle collisions \cite{Hulet}.

The gain and loss rates can be taken into account by adding absorptive terms
to the NLSE:%
\begin{equation}
i\hbar \frac{\partial \psi }{\partial t}=\left[ -\frac{\hbar ^{2}}{2m}\nabla
^{2}+V(r)-U_{0}|\psi |^{2}\right] \psi +\frac{i\hbar }{2}\left( \gamma
-2A|\psi |^{2}-3B|\psi |^{4}\right) \psi .
\end{equation}%
The initial wave function is taken to be the ground-state eigenfunction in
the harmonic potential:

\begin{equation}
\psi _{0}(r)=\pi ^{-3/4}d_{0}^{-3/2}\sqrt{N_{0}}\exp \left(
-r^{2}/d_{0}^{2}\right) ,
\end{equation}%
where $N_{0}$ is the initial number of particles. Note that the coefficients 
$A$ and $B$ were defined as rates rather than coupling constants, and do not
need corrections factors arising from Bose symmetry.

We assume that $\psi $ is spherically symmetric, and put%
\begin{align}
\tau & =\frac{1}{2}\omega t,  \nonumber \\
\rho & =\frac{r}{d_{0}},  \nonumber \\
\psi & =\sqrt{\frac{N}{4\pi }}d_{0}^{-3/2}\frac{u}{\rho }.
\end{align}%
The dimensionless NLSE for the reduced wave function $u(\rho ,\tau )$ reads 
\begin{equation}
\frac{\partial u}{\partial \tau }=i\left[ \frac{\partial ^{2}}{\partial \rho
^{2}}-\rho ^{2}+2g\frac{|u|^{2}}{\rho ^{2}}\right] u+\left[ \gamma -g\alpha 
\frac{|u|^{2}}{\rho ^{2}}-g^{2}\beta \frac{|u|^{4}}{\rho ^{4}}\right] u,
\label{NLSE1}
\end{equation}%
where%
\begin{align}
g& =\frac{|a|}{d_{0}}=4.87\times 10^{-4},  \nonumber \\
\gamma & =\frac{\gamma }{\omega }=6.18\times 10^{-3},  \nonumber \\
\alpha & =\frac{A}{2\pi \omega d_{0}|a|}=1.46\times 10^{-4},  \nonumber \\
\beta & =\frac{3B}{\left( 4\pi \right) ^{2}\omega d_{0}^{4}|a|^{2}}%
=2.616\times 10^{-5}.
\end{align}%
The instantaneous particle number, given by%
\begin{equation}
N(\tau )=\int_{0}^{\infty }d\rho |u(\rho ,\tau )|^{2},
\end{equation}%
is no longer a constant of the motion. The initial wave function is chosen
to be%
\begin{equation}
u(\rho ,0)=2\pi ^{-1/4}\sqrt{N_{0}}\rho \exp \left( -\rho ^{2}/2\right) .
\end{equation}%
The initial particle number $N_{0}=N(0)$ is the only adjustable parameter of
the model.

\section{Comparison with experiments}

We solve the modified NLSE (\ref{NLSE1}) numerically, using an algorithm due
to Goldberg {\it et al }\cite{Goldberg}{\it , }which is a version of the
popular Crank-Nicholson method \cite{Press}. The computational spatial
lattice has size $L\rho =10000,$ with spacing $d\rho =0.0008$, which
corresponds to $2.53\times 10^{-7}$ cm. The temporal grid is of size $%
L_{\tau }=40000,$ with spacing $d\tau =0.017$, which corresponds to $%
3.74\times 10^{-5}$ s. The boundary conditions are $u(0,\tau )=u(L_{\rho
},\tau )=0$.

The Rice experiment \cite{Bradley} consists of three groups of measurements
of $N(\tau )$ with different $N_{0}.$ The initial values were quoted as $%
N_{0}=500,100,0,$ with errors of $\pm 60.$ Each data point was the average
of 5 separate runs. \ We fit the data by choosing the best $N_{0}$ for each
group, and also perform an average over \ a uniform distribution of $N_{0}$
centered about the chosen value, with a width adjusted to yield the best
visual fit. The three groups of calculations correspond to the following
distributions:%
\[
\begin{array}{ccc}
\text{Group} & \left\langle N_{0}\right\rangle & \text{Distribution} \\ 
\text{I} & 325 & 385,355,325,295,265 \\ 
\text{II} & 75 & 85,80,75,70.65 \\ 
\text{III} & 31 & 61,46,31.16,1%
\end{array}%
\]

The comparison with data is shown in Fig.1. For different $N_{0}$, the
calculated $N(\tau )$ has practically the the same functional form, except
that the time origin is shifted. \ Typically, $N(\tau )$ rises exponentially
to a maximum of 1200, and then suddenly collapses to about 450. Thereafter,
it exhibits a periodic saw-tooth pattern of growth and collapse.$\ $When
averaged over a uniform distribution of $N_{0},$ the initial rise remains
unchanged, but the subsequent cycles tend to be smoothed out. The
experimental distribution was surely not uniform, but no attempts were made
to vary the distribution function to achieve a more precise fit.

The behavior of the initial rise depends mainly on the growth coefficient $%
\gamma $, and is insensitive to the loss coefficients $\alpha $ and $\beta $%
. No discernible difference is found by varying $\alpha $ and $\beta $ by
factors of 2 to 3. On the other hand, the behavior after the first collapse
is sensitive to $\alpha $, $\beta $, and to variations of the initial wave
function. Unfortunately, the data in that region is not good enough for
comparison with theory.

In our model, we have neglected quantum tunneling, the saturation of the
gain rate, trap asymmetry, and other effects. The fact that we can fit the
data indicates that these effects are of secondary importance.

\section{Details of the collapse process}

We can understand some features of the collapse process by examining the
wave function just before and after the collapse. In order to mark the time
more precisely, we show in Fig.2 a plot of $N$ as a function of time $t,$
which is measured in units of $\ 50\,d\tau =1.87\times 10^{-3}$ s. With an
initial value $N_{0}=325,$\ the first collapse happens at $t=127.$

In Fig.3, we show the reduced wave functions $u(x,t)$ for $t=100,127,128$.
The position $x$ is the site number $x$ of the computational lattice. We see
that, as time goes on, the peak of the wave function narrows as it migrates
to smaller $x,$ reaching $x=300$ at $t=127,$ when the collapse begins. At $%
t=128$ the peak has moved to $x=5$, as shown in the inset in Fig.3. The
particles under the peak are thus suddenly squeezed into a state of very
high density. This is in the region referred to as the black hole. The area
under the peak, minus background, corresponds to about 750 partilces. The
peak has disapeared by $t=130,$ as we can see in Fig.4. This indicates that
the loss mechanisms, which are proportional to high powers of the density,
became activated suddenly, and drained the black hole. Once this happens,
the density drops, and the loss mechanisms become dormant. The particle
number grows due to the gain mechanicm, and the wave function slowly
recovers an approximately Gaussian shape, as seen in Fig.4.

According to Fig.1, $1200-450=750$ particles were lost during the first
collapse. This is consistent with loss in the balck hole, and indicates that
loss mechanisms are effective only in the black hole dring a very brief time
interval. This explains why the behavior of $N(t)$ is not sensitive to the
loss coefficients. In our coarse-grained time, the interval of loss amount
to 0.004 s. Although this is very short compared to the oscillator period,
it is long compared to atomic collison times. A study at smaller time scales
would require going beyond the mean-field approxiamtion. Work in this
direction has been initiated by Saito and Ueda\cite{Saito}.

As we can see from Fig.4, after the first collapse, the wave function
acquires a persistent chaotic component. This means that the wave function
it is very sensitive to variations in the parameters of \ the model, and the
initial wave function, as has been pointed out \cite{Eleftheriou}. In Fig.5,
we examine the chaotic feature moser closely by plotting at different times
the superfluid velocity field%
\begin{equation}
v_{s}=\frac{\hbar }{m}\frac{\partial \varphi }{\partial r},
\end{equation}%
where $\varphi $ is the phase of the condensate wave function. At the onset
of \ collapse $\ $at $t=127$, the velocity is everywhere negative,
indicating a flow towards the center. It becomes oscillatory at $t=128,$ and
present a more an more chaotic look as time goes on. This suggests that the $%
^{7}$Li condensate may be a good medium for experimental and theoretical
studies of superfluid turbulence.

I thank Randy Hulet for helpful discussions. This work is support in part by
a DOE cooperative agreement DE-FC02-94ER40818.

\bigskip

{\LARGE Figure Captions}

Fig.1. Grow-collapse cycles of the particle number. Data points are solid
circles, with error bars omitted for clarity. The statistical error is
typically 20\%. Thick solid lines are predictions of the model for given $%
N_{0}$. The dotted lines indicate results of averaging over a uniform
distribution of $N_{0}$ described in the text.

Fig.2. \ Particle number as function of. time $t$ measure in units of 50
computational steps, corresponding to $1.87\times 10^{-3}$ s.

Fig.3. The reduced condensate wave function just before and just after the
collapse. The position $x$ is the site number on the computational grid,
with a spacing corresponding to $2.53\times 10^{-7}$ cm. The inset shows
behaviors near the origin, in the black hole. At $t=128,\ $there are about
750 particles under the peak at $x=5$. They are purged from the system
during the next time step, in about 0.002 s.

Fig.4. After the collapse, the loss mechanism become inactive, and the
system gains particles from the thermal cloud. The wave function recovers an
approximately Gaussian shape, but a chaotic component remains.

Fig.5. The superfluid velocity field at various times. There is a flow
towards the center just before the collapse at $t=127.$ Thereafter it looks
increasing turbulent.

\end{document}